# Mixed Reality Heritage Performance As a Decolonising Tool for Heritage Sites

Mariza Dima

Brunel University London, mariza.dima@brunel.ac.uk

Damon Daylamani-Zad

Brunel University London, damon.daylamani-zad@brunel.ac.uk

Vangelis Lympouridis

Enosis VR, vangelis@enosisvr.com

In this paper we introduce two world-first Mixed Reality (MR) experiences that fuse smart AR glasses and live theatre and take place in a heritage site with the purpose to reveal the site's hidden and difficult histories about slavery. We term these unique general audience experiences Mixed Reality Heritage Performances (MRHP). Along with the development of our initial two performances we designed and developed a tool and guidelines that can help heritage organisations with their decolonising process by critically engaging the public with under-represented voices and viewpoints of troubled European and colonial narratives. The evaluations showed the embodied and affective potential of MRHP to attract and educate heritage audiences visitors. Insights of the design process are being formulated into an extensive design toolkit that aims to support experience design, theatre and heritage professionals to collaboratively carry out similar projects.

**CCS CONCEPTS • Human-centered Computing ~ Interaction design ~ Empirical studies in interaction design • Computing Methodologies ~ Computer graphics ~ Graphics systems and interfaces ~ Mixed/augmented realit**y • **Applied Computing ~ Arts and humanities ~ Performing Arts** • Human-centered Computing ~ Interaction design ~ Interaction design process and methods

**Additional Keywords and Phrases:** theatre, real-time 3D, virtual production, Internet of Things, cultural heritage

## 1 INTRODUCTION

The heritage industry has been struggling with the legacy of colonialism for decades. However, only recently have curators and programmers actively challenged dominant colonial power structures embedded in heritage sites [1]. The process of decolonisation is not limited in the diversification of representation in exhibitions, museum staff and leadership, but extends to the interpretation perspective and the ways the public engages with the curated spaces, exhibits and experiences. Decolonisation 'concerns the proactive identification, interrogation, deconstruction and replacement of hierarchies of power that replicate colonial structures' [2]. In recent years the heritage industry has used specially designed creative experiences such as on-site installations, audio-visual media, and theatre to reframe their narratives. Immersive storytelling media such as VR experiences, AR mobile apps, and immersive heritage performances, promenade participatory theatre in the heritage site, are increasingly used to enhance engagement with the tangible and intangible heritage of the site [3].

As heritage sites have been a well researched field of application of mobile AR experiences they continue to provide a rich ground for Mixed Reality (MR) explorations inside and outside academia. The narrative and participation possibilities of immersive technologies offer an appropriate new tool for critical and affective engagement with a site's history [4] and a few projects have harnessed this potential [e.g. 5,6,7]. In this project we brought together smart glass Augmented Reality and immersive heritage performance to create a new medium for critical engagement with the site under the context of decolonisation.

The project's aim was not only to create the experiences, but also gain insights in the complex cross disciplinary and cross institutional collaboration and development process. As such, one of the main contributions of our work is the development of a design toolkit that offers a set of guidelines for professionals across creative, administrative and production departments in heritage organizations, and theatre companies on how to collaborate with immersive media specialists to design and produce specialized experiences. At the same time, it encapsulates the processes required for immersive media specialists to design experiences that empower visitors in heritage sites to discover contextual historic information and serve the set scope and broader mission of the heritage institution.

## 2 MIXED REALITY HERITAGE PERFORMANCES

'Sancho's Journey' and 'Jin's Dream' are two innovative MR experiences/performances (MRHP) designed for two heritage sites in London, UK and Deerfield, Massachusetts, US respectively. Both experiences focus on under-represented stories from 18th century enslaved and freed slave populations living in both sites while they reveal links to the transatlantic slave trade. Jointly funded by US and UK funding bodies under the scheme of New Directions for Digital Scholarship in Cultural Institutions, the experiences we created were a collaborative effort from an international team that spanned across different disciplines; interaction design, public history, theatre direction, playwright, interpretation, and curation, with many other contributors participating from certain points onwards such as digital engagement officers, volunteer coordinators, property managers, and actors. The team was consulted on decolonisation by sector professionals and worked closely with a youth panel from the International Slavery Museum, National Museums Liverpool, with experience in exhibitions and projects related to enslavement and colonialism. The panel were paid consultants in the project and their role was to challenge and inform the design of the experiences.

In both sites, visitors encountered and engaged with ideologies, beliefs, and societal frameworks of the era that gave a glimpse of how slavery was embedded and justified within society, while at the same time the performance gave voice to the experiences and cultures of enslaved people in Jamaica and British Honduras placing audiences in a reflective process. We ran over 30 performances of 'Jin's Dream' at Ashley House in Deerfield and 30 of 'Sancho's Journey' at Marble Hill House in London, both lasting for 45 minutes with approximately 80 participants in total for each site. In both performances groups of three to four visitors wear a Microsoft Hololens 2 HMD and participate in the performance as house guests at a specific day in the 1700s guided by live actors in different rooms while in specific moments holographic augmentations enhance live acting. The actors, who are not wearing HMDs, are leading the performance, and interacting with the audience and with the holographic content that the audience sees through a novel IoT sensor network and timeline system (Figures 1&2). The system is designed to make the interchange seamless and to enable the occurrence of 'magical moments' where the virtual and physical content are in synchrony blurring what is perceived as 'real'.

Actors wear an M5Core2 IoT prototyping device, under their costumes, through which they are able to trigger content in the Hololens as well as receive a haptic signal, buzz, that is a cue to continue with acting or make an acknowledgment. The devices are connected to a timeline system that controls what is rendered in the Hololens based on a set of predefined scenes and actions implemented following the narrative. In our interaction design philosophy the MR part was designed as another actor in the performance and we regarded the narrative script as one that incorporates all elements of the performance such as the dialogues, notes for the performers and all the spatial and audio augmentations happening through the MR device. In both experiences the IoT network sends actors a haptic cue to continue with their lines once the holographic content ends while in certain sections the actors control when the holographic content appears. In 'Sancho's Journey' we additionally made use of the affordances of the cyber physical space we enabled by using virtual environmental sensing such as spatial colliders to control the appearance of virtual content by both actors and audience.

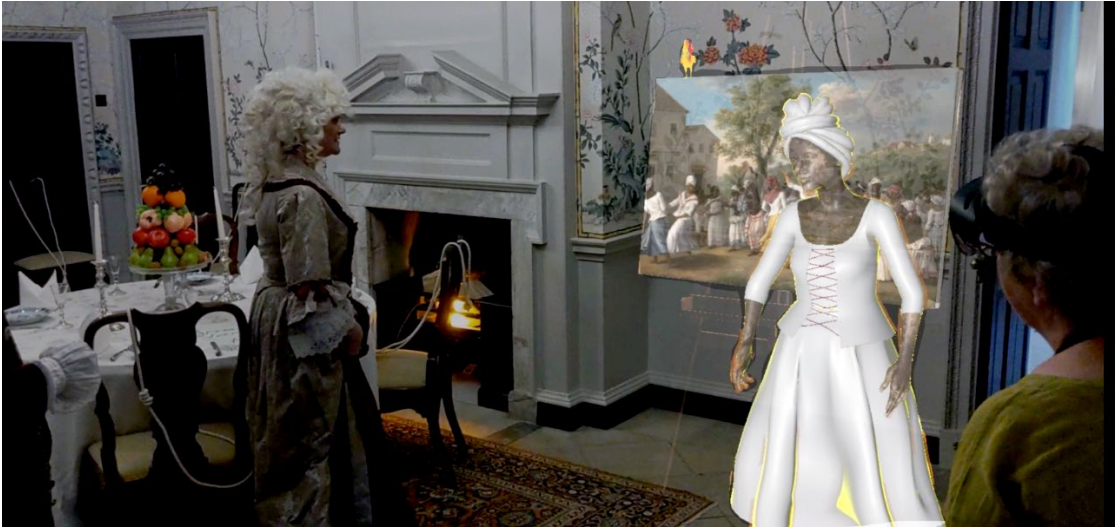

Figure 1: Actor, audience and virtual material as seen by the audience. Picture taken during 'Sancho's Journey' from an audience's Hololens HMD.

### 2.1 Evaluation Insights

The performances were evaluated for usability, engagement and learning using semi-structured interviews and in-situ observation by the actors. While the results are still being analyzed, initial insights showed a heightened sense of engagement that overcame any discomfort of wearing the HMD for a prolonged period of time. The 'magic moments' that the seamless interplay between the acting and the MR content created was mentioned by many as catalytic to feeling totally absorbed in the performance. The dramaturgical design of the experience allowed for important historical information to be transmitted in an affective way through the interaction with the actors and the engagement with the virtual content. There was a strong experiential learning component for participants of both experiences in different ways as the two sites had different kinds of connections to slavery. Ashley House visitors discovered not only the existence of enslaved Africans as servants in the houses of the 18th century Deerfield but also the extent of slavery in the northern states and their participation in an economy heavily based on the transatlantic slave trade. While there had never been enslaved people at Marble Hill, the house had financial and material connections to the transatlantic slave trade through objects such as a mahogany staircase, and tea and sugar commodities. The use of these commodities in the performance as products of enslaved labour shifted the usual focus of the stately home's visit on the then high society and histories of imperial Britain to reveal the site's entanglement with the global commodities network that existed on enslavement and exploitation. Most participants came out of the experience having learned things they did not know before, even if they were aware of or even had good knowledge of the site's connection to slavery.

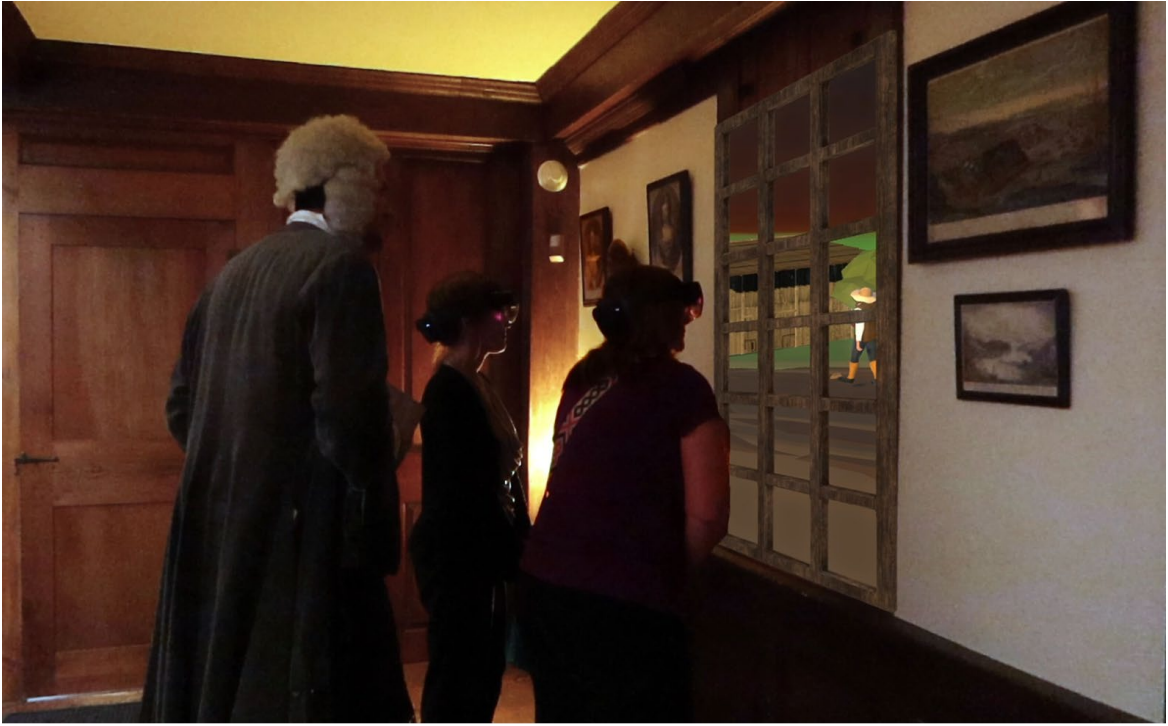

Figure 2: Actor, audience and virtual material as seen by the audience. Picture taken during 'Jin's Dream' from an audience's Hololens HMD.

In terms of setting the experience, evaluation offered valuable insights on how to create useful on-boarding and off-boarding experiences as well as how to integrate them within the performance. One of the most interesting points raised by participants was a desire to participate in the dialogues more actively, and some of them indeed did, however, they emphasised that they were not certain at which point they could do so before any holographic audiovisual material appeared in their HMDs. This observation opens an interesting discussion on participant agency in MRHP.

## 3   DESIGN TOOLKIT

Creating the two experiences was a complex collaborative process between different disciplines and groups, and required the development of an appropriate co-creation model, a unique pipeline system for production, as well as a framework for interaction design of the experience. We have reflected on our process and abstracted it into a design toolkit that can be used by experience design, theatre and heritage professionals to carry out similar projects. In fact, as the toolkit outlines a model for collaborative creation of immersive experiences, it is useful for a plethora of cases where MR or other immersive technologies are applied in several contexts. The toolkit is at the latest stages of development and will be soon available for free.

## ACKNOWLEDGMENTS

This work was supported by a joint **Arts and Humanities Research Council** (UK, AH/W005530/1) and **National Endowment For the Humanities** (US, HND-284975-22) research grant under the theme New Direction for Digital Scholarship in Cultural Institutions.

## REFERENCES

[1] Maggie Appleton. 2020. Empowering Collections 2030, Museum Association, Retrieved March 7, 2024 from: https://media.museumsassociation.org/app/uploads/2020/06/11085829/MS1681-Empowering-collections__v8.pdf
[2] John Giblin, Imma Ramos, and Nikki Grout. 2019. Dismantling the Master's House, Third Text, 33:4-5, 471-486
[3] Jenny Kidd. 2018 Immersive' Heritage Encounters. The Museum Review, 3(1), Retrieved March 7, 2024 from: http://articles.themuseumreview.org/tmr_vol3no1_kidd
[4] Alyssa K Loh. 2017. I Feel You. Artforum. Retrieved March 7, 2024 from: https://www.artforum.com/print/201709/alyssa-k-loh-on-virtual-reality-and-empathy-71781
[5] Mariza Dima. 2022. A Design Framework for Smart Glass Augmented Reality Experiences in Heritage Sites. J. Comput. Cult. Herit. 15, 4, Article 66 (December 2022), 19 pages. https://doi.org/10.1145/3490393
[6] Fabula. 2018. Bitter Wind: Greek tragedy for Hololens. Retrieved March 7, 2024 from https://sites.wustl.edu/fabulab/bitter-wind-greek-tragedy-for-hololens/
[7] Tanya Krzywinska, Tim Phillips, Alcwyn Parker, and Michael James Scott. 2020. From Immersion's Bleeding Edge to the Augmented Telegrapher: A Method for Creating Mixed Reality Games for Museum and Heritage Contexts. ACM Journal of Computing and cultural Heritage, 13,4: Article 0.Claudia M. Leue, Timothy Jung, Dario tom Dieck. 2015. Google Glass Augmented Reality: Generic Learning Outcomes for Art Galleries. In: Tussyadiah I., Inversini A. (eds) Information and Communication Technologies in Tourism 2015. Springer, Cham. https://doi.org/10.1007/978-3-319-14343-9_34.